\documentclass{vldb}
\usepackage{balance}
\usepackage[ruled,vlined,linesnumbered]{algorithm2e}
\usepackage{url}
\usepackage{color}
\usepackage{subfigure}
\newdef{definition}{Definition}

\newtheorem{example}{Example}

\begin{document}
%


\title{Cohort Query Processing}

\numberofauthors{1}
\author{
	\alignauthor
	Dawei Jiang$^{\dag}$ \hspace{1em} Qingchao Cai$^{\ddag}$ \hspace{1em} Gang Chen$^{\dag}$ \hspace{1em} H. V. Jagadish$^{\S}$ \\
	\hspace{1em} Beng Chin Ooi$^{\ddag}$ \hspace{1em}   Kian-Lee Tan$^{\ddag}$ \hspace{1em} Anthony K. H. Tung$^{\ddag}$\\
	\vspace{.1cm}
	\affaddr{$^\dag$Zhejiang University \qquad $^\ddag$National University of Singapore \qquad $^\S$University of Michigan}\\
    \email{\mbox{\hspace{-1em} $^\dag$\{jiangdw, cg\}@zju.edu.cn $^{\ddag}$\{caiqc, ooibc, tankl, atung\}@comp.nus.edu.sg  $^\S$jag@umich.edu}}
}
\maketitle

\begin{abstract}
    Modern Internet applications often produce a large volume of user activity
    records.  Data analysts are interested in cohort analysis, or finding
    unusual user behavioral trends, in these large tables of activity records.
    In a traditional database system, cohort analysis queries are both painful
    to specify and expensive to evaluate.  We propose to extend database systems
    to support cohort analysis. We do so by extending SQL with three new
    operators. We devise three different evaluation schemes for cohort query processing.
    Two of them adopt a non-intrusive approach. The third approach employs
    a columnar based evaluation scheme with optimizations specifically
    designed for cohort query processing.
    Our experimental results confirm
    the performance benefits of our proposed columnar database system, compared
    against the two non-intrusive approaches that implement cohort queries on top of regular
    relational databases.

\end{abstract}




\SetKwInOut{Input}{Input}
\SetKwProg{Fn}{}{}{}
\SetKwFunction{GetBirthTuple}{GetBirthTuple}
\SetKwFunction{Open}{Open}
\SetKwFunction{GetCurrent}{GetCurrent}
\SetKwFunction{GetNext}{GetNext}
\SetKwFunction{GetNextUser}{GetNextUser}
\SetKwFunction{SkipCurUser}{SkipCurUser}
\SetKwFunction{Close}{Close}
\SetKwFunction{Birth}{Birth}
\DontPrintSemicolon

\section{Introduction}
\label{sec1}

\begin{table*}
\centering
\makebox[0pt][c]{\parbox{1.0\textwidth}{%
    \begin{minipage}[t]{0.7\hsize}\centering
        \caption{Mobile Game Activity Table}
        \label{evtable}
        \small
        \begin{tabular}{|l||llllll|}
            \hline
            &player & time & action & role & country & gold \\\hline
            $t_1$ &001 & 2013/05/19:1000 & launch & dwarf & Australia & 0 \\
            $t_2$ &001 & 2013/05/20:0800 & shop & dwarf & Australia & 50 \\
            $t_3$ &001 & 2013/05/20:1400 & shop & dwarf & Australia & 100 \\
            $t_4$ &001 & 2013/05/21:1400 & shop & assassin & Australia & 50 \\
            $t_5$ &001 & 2013/05/22:0900 & fight & assassin & Australia & 0 \\
            $t_6$ &002 & 2013/05/20:0900 & launch & wizard & United States & 0 \\
            $t_7$ &002 & 2013/05/21:1500 & shop & wizard & United States & 30 \\
            $t_8$ &002 & 2013/05/22:1700 & shop & wizard & United States & 40 \\
            $t_9$ &003 & 2013/05/20:1000 & launch & bandit & China & 0 \\
            $t_{10}$ &003 & 2013/05/21:1000 & fight & bandit & China & 0 \\
            \hline
        \end{tabular}
    \end{minipage}%
    \begin{minipage}[t]{0.2\hsize}\centering
        \caption{Results of $Q_s$}
        \label{olapreport}
        \small
        \begin{tabular}{|ll|}
            \hline
            week &  avgSpent \\ \hline
            2013-05-19 & 50 \\
            2013-05-26 & 45 \\
            2013-06-02 & 43 \\
            2013-06-09 & 42 \\
            2013-06-16 & 45 \\
            \hline
        \end{tabular}
    \end{minipage}
}}
\end{table*}

Internet applications often accumulate a huge amount of
activity data representing information associated with user
actions. Such activity
data are often tabulated to provide insights into the
behavior of users in order to increase sales and ensure
user retention. To illustrate, Table \ref{evtable} shows some samples
of a real dataset containing user activities
collected from a mobile game. Each tuple in this table represents
a user action and its associated information. For example, tuple $t_1$
represents that player 001 launched the game on 2013/05/19 in Australia
in a {\tt dwarf} role.

An obvious solution to obtain insights from such activity data is to apply
traditional SQL {\tt GROUP BY} operators. For example,
if we want to look at the players' shopping
trend in terms of the {\tt gold} (the virtual currency) they spent, we may run
the following SQL query $Q_s$.
\begin{tabbing}
\small
\texttt{\small SELECT week, Avg(gold) as avgSpent}\\
\texttt{\small FROM GameActions} \\
\texttt{\small WHERE action = "shop"} \\
\texttt{\small GROUP BY Week(time) as week}
\end{tabbing}

Executing this query against a sample dataset (of which Table \ref{evtable} shows some records) results in Table
\ref{olapreport},
where each tuple represents
the average gold that users spent each week.
The results seem to suggest
that there was a slight drop in shopping, and then a partial recovery. However,
it is hard to draw meaningful
insights.

There are two major sources that
can affect human behavior \cite{cohortbook}: 1) aging, i.e.,
people behave differently as they grow older and
2) social changes, i.e., people behave differently if the societies they live in are different.
In our in-game shopping example, players tend to buy more weapons
in their initial game sessions
than they do in later game sessions - this is the effect of aging.
On the other hand, social change may also affect the players' shopping
behavior, e.g., with new weapons being introduced in iterative game
development, players may start to spend again in order to
acquire these weapons. \textbf{Cohort analysis},
originally introduced in Social Science, is a data analytical technique
for assessing the effects of aging on human behavior in a changing society \cite{cohortbook}.
In particular, it allows us to tease apart the effect of aging
from the effect of social change, and hence can offer more valuable
insights.

With cohort analytics, social scientists study the human behavioral trend
in three steps: 1) group users into cohorts; 2) determine the age of user 
activities and 3) compute aggregations for each (cohort, age) bucket.
The first step employs the so called {\tt cohort} operation
to capture the effect of social differences.
Social scientists choose a particular action $e$ (called the birth action) and
divide users into different groups (called cohorts) based on the
first time (called birth time) that users performed $e$. Each cohort is then represented by the
time bin (e.g., day, week or month) associated with the birth time \footnote{The interval of
the time bin is chosen to ensure that there are no significant social differences
occurred in that time bin.}.
Each activity tuple of a user is then assigned to the same cohort that this user belongs to.
In the second step, social scientists capture the effect of aging by partitioning
activity tuples in each cohort into smaller
sub-partitions based on \textbf{age}. The age of an activity tuple $t$ is the duration
between the birth time of that user and
the time that user performed $t$. Finally, aggregated behavioral measure
is reported for each (cohort, age) bucket.

\begin{figure}[tb]
  \centering
  \includegraphics[width=.22\textwidth]{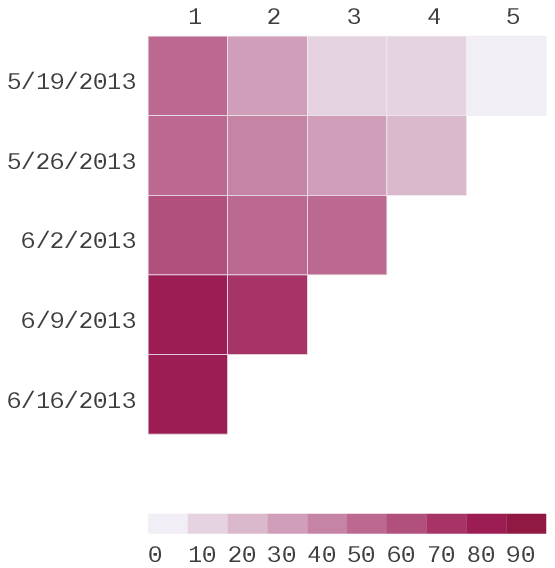}\\
  \caption{Launch Cohort Shopping Trend}\label{fig3}
  \vspace{-3mm}
\end{figure}

Back to our in-game shopping example, suppose we choose \texttt{launch} as the birth action
and week as the cohort time bin interval,
the activity tuples of player 001 are assigned to 2013-05-19
launch cohort since the activity tuple $t_1$ (called birth activity tuple)
indicates that player 001 first launched the game at
that week. We further partition activity tuples in 2013-05-19 launch cohort
into sub-partitions identified by age.
As an example, the activity tuple $t_2$ is partitioned into the week 1 age sub-partition.
Finally, we report the average gold spent for each (cohort, age) bucket shown in Table \ref{cohortreport}
and visualized in Figure \ref{fig3}. Each row in Figure \ref{fig3} represents
the shopping trend of a cohort measured at different ages.
For example, column $k$ (under the age section)
reports the average expenditures of the cohorts
at the $k$th week (i.e., age = $k$).

\begin{table}[tb]
\small
\centering
\caption{Cohort Report for Shopping Trend}
\label{cohortreport}
\begin{tabular}{|llllll|}
\hline
cohort & \multicolumn{5}{c|}{age (weeks)} \\ \hline
& 1 & 2 & 3 & 4 & 5 \\ \hline
2013-05-19 (145) & 52 & 31 & 18 & 12 & 5 \\
2013-05-26 (130) & 58 & 43 & 31 & 21 & \\
2013-06-02 (135) & 68 & 58 & 50 &  &  \\
2013-06-09 (140) & 80 & 73 & & & \\
2013-06-16 (126) & 86 & & & & \\
\hline
\end{tabular}
\end{table}

By looking at each row horizontally, we can see the aging effect,
i.e., players spent more gold on buying weapons on their initial game
sessions than their later game sessions.
For example, for the {\tt 2013-05-19} cohort, the amount of gold spent
was 52 in the first week, but there is a clear declining trend as
time passes.
On the other hand,
by comparing different rows (i.e., reading rows vertically), we can observe
that the drop-off trend seems to becomes less severe.
Comparing rows 1 and 2 in our example, we observe that for the same column,
the value of row 2 is larger than that of row 1.
This suggests that the iterative game development indeed did a better job
of retaining player enthusiasm as they aged, an insight which cannot be
drawn from OLAP style results in Table \ref{olapreport}.

While the classic cohort analysis we presented so far has been
very useful in many analytic applications such as retention analysis
\cite{mixpanel}, we hope to 1) generalize the approach so that it can
be applied to a wider range of applications and 2) extend SQL
to support this generalized cohort analysis
in this paper.

There are two limitations in the standard social science style
cohort analysis. First, social scientists typically analyze
a whole dataset. This is because the datasets used are usually small
and are specifically collected for a certain cohort analysis task.
As such, there is no mechanism for extracting
a portion of users or activity tuples for cohort analysis.
While this task seems trivial, it has to be handled with care
as a naive selection may lead to incorrect answers!
Referring to our running example (from Table \ref{evtable}),
suppose we choose \texttt{launch}
as the birth action, and we are interested to perform a cohort analysis
on those tuples {\tt where time > 2013/05/22:0000}.
Now, the resultant subset of tuples
is $\{t_5, t_8\}$. However, we no longer can perform any cohort analysis
as the birth activity tuple $t_1$, i.e., the activity
tuple representing player 001 first performing \texttt{launch}, has
been removed.
Second, social scientists use only the time attribute to identify cohorts.
This is because time is considered to be the key attribute
that determines social change. However, it is possible that some other
attributes, such as country, may also have a significant effect on
social differences, in which case, it may be interesting to
perform cohorts based on the country attribute. Thus, it would be desirable
to provide support for a more general cohort analysis task.

\begin{figure*}[tbh]
	\label{fig:translation}
\centering
\small
\subfigure[]{
\label{subfig:a}
        \begin{minipage}[t]{0.28\textwidth}
        \begin{tabbing}
        {\tt WITH birth AS(} \\
        \quad {\tt SELECT p, Min(t) as birthTime} \\
        \quad {\tt FROM $D$}\\
        \quad {\tt WHERE a = "launch"}\\
        \quad {\tt GROUP BY p} \\
        {\tt ),}
        \end{tabbing}
        \end{minipage}
}
\subfigure[]{
\label{subfig:b}
        \begin{minipage}[t]{0.3\textwidth}
        \begin{tabbing}
        {\tt birthTuples AS (}\\
        \quad{\tt SELECT p, c as cohort, birthTime}\\
        \quad\quad\quad\quad\ \ {\tt role as birthRole}\\
        \quad{\tt FROM $D$, birth}\\
        \quad{\tt WHERE $D$.p = birth.p AND}\\
        \quad\quad\quad\quad{\tt $D$.t = birth.birthTime),}
        \end{tabbing}
        \end{minipage}
}
\subfigure[]{
\label{subfig:c}
    \begin{minipage}[t]{0.32\textwidth}
    \begin{tabbing}
    {\tt cohortT AS (}\\
    \quad{\tt SELECT p, a, cohort, birthRole, gold} \\
    \quad\quad\quad\quad\ {\tt TimeDiff($D$.t, birthTime) as age} \\
    \quad\texttt{FROM $D$, birthTuples} \\
    \quad\texttt{WHERE $D$.p = birthTuples.p),}
    \end{tabbing}
    \end{minipage}
}\\
\subfigure[] {
\label{subfig:d}
    \begin{minipage}[t]{0.5\textwidth}
    \begin{tabbing}
    \texttt{cohortSize AS (}\\
    \quad\texttt{SELECT cohort, Count(distinct p) as size}\\
    \quad\texttt{FROM cohortT} \\
    \quad\texttt{GROUP BY cohort}\\
    {\tt ),}
    \end{tabbing}
    \end{minipage}
}
\subfigure[] {
\label{subfig:outer}
    \begin{minipage}[t]{0.45\textwidth}
    \begin{tabbing}
    \texttt{SELECT cohort, size, age, Sum(gold)}\\
    \texttt{FROM cohortT, cohortSize}\\
    \texttt{WHERE cohortT.cohort = cohortSize.cohort} \\
    \quad\quad\quad\texttt{birthRole = "dwarf" AND a = "shop" AND age > 0} \\
    \texttt{GROUP BY cohort, age}
    \end{tabbing}
    \end{minipage}
}
\caption{The SQL query $Q_s$ of our example analysis task.}\label{fig:sqlofq1}
\end{figure*}

In this paper, we make the following contributions to address the above issues.
\begin{itemize}
\item We define the important problem of cohort analytics in the context of a DBMS.
\vspace{-1mm}
\item We introduce an extended relation to model user activity data for cohort analytics,
and introduce three new operators for manipulating the extended
relation and composing cohort queries.
Two of the operators are designed for extracting a subset of activities for cohort analysis,
and the last one is designed for producing aggregates over arbitrary attribute combinations.
We show that cohort queries can be expressed elegantly and concisely using the data model and the newly proposed operators.
We also show how
more complicated data analytics tasks can be expressed
using a mix of traditional SQL clauses and the newly proposed operators.
\vspace{-1mm}
\item We build a columnar based cohort query engine, COHAHA, which
implements multiple optimizations for efficient cohort query processing.
\vspace{-1mm}
\item We design a benchmark study to compare
COHANA against alternative non-intrusive evaluation schemes in terms of cohort
query processing performance.
The experimental results show that COHANA is two orders superior to its
mostly optimized counterpart, demonstrating the necessity of
extending database systems to cater to cohort analytics,
rather than simply running an SQL statement over
a traditional DBMS.
\end{itemize}

The rest of the paper is organized as follows:
Section \ref{sec3} presents the SQL based and the materialized view based approaches
for processing cohort queries.
Section \ref{sec2} presents the foundations of cohort analysis.
Section \ref{sec4} presents our proposed columnar database based scheme
for cohort query processing.
Section \ref{sec5} reports the experimental results. We present related work in Section \ref{sec6}
and conclude 
in Section \ref{sec7}.

\section{Non-intrusive Approaches to Cohort Analytics}
\label{sec3}

A least intrusive approach to supporting cohort analytics
is to use an existing relational DBMS and 
express the cohort analysis task as a SQL query.
%
%
%
%
We illustrate such an approach using the following cohort analysis task:
\begin{example}
\label{example1}
Given the {\tt launch} birth action and the activity table as shown
in Table~\ref{evtable} (denoted as $D$), for players who play the
{\tt dwarf} role at their birth time, cohort those players based on
their birth countries and
report the total gold that country launch cohorts spent
since they were born.
\end{example}

Figure~\ref{fig:sqlofq1} shows the corresponding SQL query $Q_s$ for this
task.
To save space, we use {\tt p, a, t, c} abbreviations respectively to denote
the {\tt player}, {\tt action}, {\tt time}, and {\tt country} 
attribute in Table \ref{evtable}.
The $Q_s$ employs four sub-queries (i.e., Figure~\ref{subfig:a} -- Figure~\ref{subfig:d})
and one outer query (i.e., Figure~\ref{subfig:outer}) to produce the results.
Overall, this SQL approach performs poorly for three reasons:
\begin{itemize}
  \item The SQL statement $Q_s$ is verbose,
	  and its complexity renders it prone to mistakes.
  \item The SQL statement $Q_s$ requires many joins to perform the analysis task.
	  As we shall see in our experimental study, such a query processing scheme
	  can be up to 5 orders of magnitude slower than our proposed solution.
  \item It requires manual tuning.  For example, one may notice that we can push the selection condition
  (i.e., \texttt{birthRole = "dwarf"}) from the outer query (Figure~\ref{subfig:outer})
  to the inner sub-query (Figure~\ref{subfig:c}) to reduce the size of intermediate
  tables. Ideally, such an optimization can be performed by an intelligent optimizer.
  However, our evaluation shows that few database systems can perform
  such an optimization.
\end{itemize}


To speed up the processing of the analysis task, we can
adopt a {\em materialized view} (MV) approach that stores some
intermediate results.
For example, we can materialize the intermediate table \texttt{cohortT}
produced by the sub-query in $Q_s$ (Figure~\ref{subfig:c})
as follows.

\begin{tabbing}
{\tt \small CREATE VIEW MATERIALIZED cohorts as cohortT}
\end{tabbing}

With \texttt{cohorts}, we can express the query $Q_S$ in a simpler
SQL statement consisting
of a single sub-query (Figure~\ref{subfig:d}) and an outer query
(Figure~\ref{subfig:outer}).
The performance of the resulting SQL expression is also
improved since it only involves a single
join. However, the materialized view approach also suffers from
a number of issues.

\begin{itemize}
\item The cost of generating the MV is still high since
    it involves two joins (Figure~\ref{subfig:b} and \ref{subfig:c}).
    \vspace{-1mm}
\item The storage space for the MV is huge if the approach is used as a general
    cohort query processing strategy. Figure~\ref{subfig:c} only includes a
    single calculated birth attribute {\tt birthRole} as it is the only attribute appearing in
    the birth selection condition (i.e., the condition of playing as the {\tt dwarf}
    role at birth time) of the analysis task. However, if other
    calculated birth attributes are
    also involved in the birth selection condition, we need to include those
    attributes in the MV as well. In the extreme case, every possible birth
    attribute shall be included in the MV, doubling the storage space
    as compared to the original activity table.
    \vspace{-1mm}
\item The MV only answers cohort queries introduced by {\tt launch} birth
    action. If another birth action (e.g., {\tt shop}) is used, one more MV is required.
    Obviously, this per birth action per MV approach does not scale even
    for a small number of
    birth actions due to the cost of MV construction and maintenance.
    \vspace{-1mm}
\item The query performance is still not optimal. By the definition of the analysis
    task, if a player does not play as {\tt dwarf} role when that player is born,
    we should exclude all activity tuples of that player in the result set.
    Ideally, this filtering operation can be performed by simply checking
    the single birth activity tuple of that player. If the birth activity tuple
    indicates that the player is not qualified, we can safely skip all activity tuples of
    that player without further checking. However, as shown in Figure~\ref{subfig:outer},
    the MV approach needs to, unnecessarily, check each activity tuple
    of a player to perform the filtering operation (i.e., comparing
    the value in {\tt birthRole} attribute against {\tt dwarf}).
    Building an index on the {\tt birthRole} attribute cannot improve the
    situation much since index look up will introduce too many random seeks
    on large activity tables.

\end{itemize}

\section{Cohort Analysis Foundations}
\label{sec2}
In this paper, we seek to
extend an existing relational
database system to support cohort analytics.
This section presents the data model, which includes a central new concept of an {\textit activity table}, and the proposed new cohort
operators.

We use the term \emph{cohort} to refer to a number of individuals
who have some common characteristic in
performing a particular action for the first time;
we use this particular action and the attribute values of
the common characteristics to identify the resulting cohort.
For example, a group of users who first {\tt login}
(the particular action) in {\tt 2015 January} (the common characteristic)
is called the {\tt 2015 January login cohort}.
Similarly customers
who make their first purchase in the United States form a
United States purchase cohort. Broadly speaking, cohort analysis is a data
exploration technique that
examines longitudinal behavioral trends
of different cohorts
since they were {\bf born}.

\subsection{Data Model}
\label{subsec:datamodeling}

We represent a collection of activity data as an instance of an \emph{activity relation},
a special relation where each tuple represents the information
associated with a single user activity. We will also call an activity relation
an activity table. In this paper, the two terms, i.e., activity relation and
activity table are used interchangeably.

Formally, an activity table $D$ is a relation with attributes
$A_u, A_t, A_e, A_1, \ldots, A_n$ where $n \geq 1$. $A_u$ is a string uniquely
identifying a user; $A_e$ is also a string,
representing an action chosen from a pre-defined collection of actions,
and $A_t$ records the time at which $A_u$ performed $A_e$.
Every other
attribute in $D$ is a standard relational attribute.
Furthermore,
an activity table has a primary key constraint on
$(A_u, A_t, A_e)$. That is,
each user $i$ can only perform a specific action $e$ once at each time instant.
As exemplified in Table \ref{evtable},
the first three columns correspond to the user ($A_u$), timestamp ($A_t$) and
action ($A_e$) attribute, respectively.
Role and Country are dimension attributes, which respectively specify the role
and the country of player $A_u$ when performing $A_e$ at $A_t$.
Following the two dimension attributes is gold, a measure attribute representing
the virtual currency that player $A_u$ spent for this action.
We shall continue to use Table \ref{evtable}
as our running example for describing each concept in cohort analysis.

\subsection{Basic Concepts of Cohort Analysis}
\label{ssec:concepts}

We present three core concepts of cohort analysis: {\bf birth action}, {\bf birth time}
and {\bf age}. Given an action
$e\in \operatorname{Dom}(A_e)$, the birth time of user $i$ is the
first time that $i$ performed $e$ or -1 if $i$ never performed $e$,
as shown in Definition \ref{def:birth}.
An action $e$ is called a birth action if $e$ is used to define
the birth time of users.

\begin{definition}
\label{def:birth}
Given an activity table $D$,
and a birth action $e\in \operatorname{Dom}(A_e)$,
a time value $t^{i,e}$ is called the birth time
of user $i$ if and only if
\begin{displaymath}
t^{i,e}=\left\{
          \begin{array}{ll}
            \min\pi_{A_t}(\sigma_{A_u = i \wedge A_e = e}(D)) & if\quad\sigma_{A_u = i \wedge A_e = e}(D) \neq \emptyset\\
            -1 & otherwise
          \end{array}
   \right.
\end{displaymath}
where $\pi$ and $\sigma$ are the standard projection and selection operators.
\end{definition}

\begin{definition}
Given an activity table $D$,
and a birth action $e\in \operatorname{Dom}(A_e)$,
a tuple $d^{i,e}\in D$ is
called the birth activity tuple of user $i$ if and only if
$$
d^{i,e}[A_u] = i \wedge d^{i,e}[A_t] = t^{i, e}
$$
\end{definition}

Since $(A_u, A_t, A_e)$ is the primary key of $D$, we conclude
that for each user $i$, there is only one birth activity tuple of $i$ in $D$
for any birth action $e$ that $i$ performed.

\begin{definition}
Given the birth time $t^{i,e}$, a numerical value $g$ is called the age of user
$i$ in tuple $d\in D$, if and only if
$$d[A_u] = i \wedge t^{i,e} >= 0 \wedge g = d[A_t] - t^{i,e}$$
\end{definition}

The concept of age is designed for specifying the time point to aggregate
the behavioral metric of a cohort. In cohort analysis, we calculate the
metric only at positive ages. That is, if the age of an user in a tuple is negative,
that tuple will be excluded from the final report.
The activity tuple with a positive
age is called an age activity tuple. Furthermore, in practical
applications, the age $g$ is normalized by a certain time unit such as a day,
week or month. Without loss of generality, we assume that the
granularity of $g$ is a day.

Consider the example activity relation in Table \ref{evtable}.
Suppose we use the action {\tt launch} as the birth action.
Then, the activity tuple $t_1$ is
the birth activity tuple of player 001, and the birth time is 2013/05/19:1000.
The activity tuple $t_2$ is an age tuple of player 001 produced at
age 1.

\subsection{Cohort Operators}
\label{subsec:cohortoperation}

We now present operations on a single activity table.
In particular, we propose two new operators to retrieve a subset of
activity tuples
for cohort analysis.
We also propose a cohort aggregation operator
for aggregating activity tuples for each (cohort, age) combination.
As we shall see, these three operators enable us to express
a cohort analysis task in a very concise and elegant way that is
easy to understand.

\subsubsection{The $\sigma^b_{C,e}$ Operator}

The birth selection operator $\sigma^b_{C,e}$ is used to retrieve activity
tuples of qualified users whose birth activity tuples satisfy a specific condition $C$.

\begin{definition}
Given an activity table $D$, the birth selection operator $\sigma^b_{C,e}$ is defined as
\begin{align*}
  \sigma^b_{C,e}(D) = \{d\in D \,|\, i \leftarrow d[A_u] \wedge C(d^{i,e}) = \text{true}\}
\end{align*}
where $C$ is a propositional formula and $e$ is a birth action.
\label{def:age}
\end{definition}

Consider the activity relation $D$ in Table \ref{evtable}. Suppose we want to
derive an activity table from $D$ which retains all activity tuples of
users who were born from performing \texttt{launch} action
in Australia. This can be achieved with the following expression,
which returns $\{t_1, t_2, t_3, t_4, t_5\}$.
$$\sigma^b_{\text{country}=\text{Australia}, \text{launch}}(D)$$

\subsubsection{The $\sigma^g_{C,e}$ Operator}

The age selection operator $\sigma^g_{C,e}$ is used to generate an activity table from $D$
which retains all birth activity tuples in $D$ but only
a subset of age activity tuples which satisfy a condition $C$.

\begin{definition}
Given an activity table $D$, the age selection operator $\sigma^g_{C,e}$
is defined as
\begin{align*}
  \sigma^g_{C,e}(D) = & \{d\in D | i \leftarrow d[A_u] \wedge \\
                      & ((d[A_t] = t^{i,e}) \vee (d[A_t] > t^{i,e} \wedge C(d) =
                      \text{true})) \}
\end{align*}
where $C$ is a propositional formula and $e$ is a birth action.
\end{definition}

For example, suppose {\tt shop} is the birth action,
and we want to derive an activity table which retains
all birth activity tuples in Table \ref{evtable}
but only includes age activity tuples which indicate users performing
in-game shopping in all countries but China. The following expression can
be used to obtain the desired activity table.
$$\sigma^g_{\text{action}=\text{shop}\wedge\text{country}\neq\text{China},\text{shop}}(D)$$
The result set of the above selection operation is $\{t_2, t_3, t_4, t_7, t_8\}$
where $t_2$ is the birth activity tuple of player 001, $t_3$ and $t_4$ are
the qualified age activity tuples of player 001. The activity tuples
$t_7$ and $t_8$
are the birth activity tuple and the qualified age activity tuple of player 002.

A common requirement in specifying $\sigma^g_{C,e}$ operation is that
we often want to reference the attribute values of birth activity tuples
in $C$. For example, given the birth action
{\tt shop}, we may want to select age activity tuples
whose users perform in-game shopping at the same location as their
country of birth.  We introduce
a \Birth{} function for this purpose. Given an attribute
$A$, for any activity tuple $d$, the \Birth{$A$} returns the value of
attribute $A$ in $d[A_u]$'s birth activity tuple:
$$\Birth{$A$} = d^{i,e}[A]$$
where $i=d[A_u]$ and $e$ is the birth action.

In our running example, suppose \texttt{shop} is the birth action, and we want
to obtain an activity table which retains all
birth activity tuples but only include
age activity tuples which indicate that players performed shopping in the same role
as they were born.
The following expression is used to retrieve the desired results.
$$\sigma^g_{\text{role}=\texttt{Birth}(\text{role}),\text{shop}}(D)$$
The result set of the above operation is $\{t_2, t_3, t_7, t_8\}$ where
$t_2$ and $t_7$ are the birth activity tuples of player 001 and player 002, respectively,
and $t_3$ and $t_8$ are the qualified age activity tuples.

\subsubsection{The $\gamma^c_{\mathcal{L},e,f_A}$ Operator}

We now present the cohort aggregation operator $\gamma^c_{\mathcal{L},e,f_A}$.
This operator produces cohort aggregates in two steps: 1) cohort users and 2)
aggregate activity tuples.

In the first step, given an activity table $D$ with its attribute set $\mathcal{A}$
and a birth action $e$, we pick up a cohort attribute set $\mathcal{L}\subset\mathcal{A}$
such that $\mathcal{L}\cap\{A_u, A_e\} = \emptyset$ and assign each user $i$
to a cohort $c$ specified by $d^{i,e}[\mathcal{L}]$. Essentially, we divide users
into cohorts based on the projection of users' birth activity tuples onto a specified
cohort attribute set.

In our running example, suppose {\tt launch} is the birth action and the cohort
attribute set is $\mathcal{L}$=\{country\}, player 001 in Table \ref{evtable}
is assigned to the Australia launch cohort, player 002 is assigned
to the United States launch cohort and player 003 is assigned to the China
launch cohort.

After assigning users to cohorts, in the second step, we divide activity tuples
(including both the birth activity tuples and the age activity tuples) into
the same cohorts that the user belonged to for aggregation.

\begin{definition}
\label{def:cohort}
Given an activity table $D$, the cohort aggregation operator
$\gamma^c_{\mathcal{L},e,f_A}$ is defined as
\begin{align*}
  \gamma^c_{\mathcal{L},e,f_A}(D) = & \{(d_{\mathcal{L}}, g, s, m) | \\
    & D_g \leftarrow \{(d, l, g) | d \in D \wedge i \leftarrow d[A_u] \\
    &  \qquad \qquad \wedge l = d^{i,e}[\mathcal{L}] \wedge g = d[A_t] - t^{i,e}  \} \\
    & \wedge (d_{\mathcal{L}}, g) \in \pi_{l, g}(D_g) \\
    & \wedge s = \text{Count}(\pi_{A_u}\sigma_{d_g[{l}]=d_{\mathcal{L}}}(D_g)) \\
    & \wedge m = f_A(\sigma_{d_g[{l}] = d_{\mathcal{L}} \wedge d_g[g] = g \wedge g > 0}(D_g))
\end{align*}
where $\mathcal{L}$ is a cohort attributes set, $e$ is a birth action
and $f_A$ is a standard aggregation function with respect to the attribute $A$.
\end{definition}

In summary, the cohort aggregation operator takes an activity table $D$
as input and produces a normal relational table $R$ as output.
Each row in the output table $R$ consists of four parts $(d_{\mathcal{L}}, g, s,
m)$, where $d_{\mathcal{L}}$ is the projection of birth activity tuples onto
the cohort attributes set $\mathcal{L}$ and identifies the cohort,
$g$ is the age, i.e., the time point that we report the aggregates,
$s$ is the size of the cohort, i.e., the number of users in the cohort specified by
$d_{\mathcal{L}}$, and $m$ is the aggregated measure produced by the aggregate
function $f_A$. Note that we only apply $f_A$ on age activity tuples with $g>0$.

\subsubsection{Properties of Cohort Operators}

We note that the two selection operators,
$\sigma^b_{C,e}$ and $\sigma^g_{C,e}$, are commutative
if they involve the same birth action.
\begin{equation}
\label{cohortlaw}
\sigma^b_{C,e}\sigma^g_{C,e}(D) = \sigma^g_{C,e}\sigma^b_{C,e}(D)
\end{equation}
Based on this property, we can, as we shall see in Section \ref{sec4},
push the birth selection operator down the query plan to
optimize cohort query evaluation.

\subsection{The Cohort Query}
\label{subsec:cohortquery}

Given an activity table $D$ and operators
$\sigma^b_{C,e}$, $\sigma^g_{C,e}$, $\pi_{\mathcal{L}}$,
and $\gamma^c_{\mathcal{L}, e, f_A}$, a cohort query $Q: D \rightarrow R$ can be
expressed as a composition of those operators that takes $D$ as input and produces
a relation $R$ as output with the constraint that the same birth action
$e$ is used for all cohort operators in $Q$. To specify a cohort query, we
propose to use the following SQL-style {\tt SELECT} statement.

\begin{tabbing}
{\small \tt SELECT ... FROM $D$}\\
{\small \tt BIRTH FROM action = $e$ [ AND $\sigma^b_{C,e}$ ]}\\
{\small \tt [ AGE ACTIVITIES IN $\sigma^g_{C,e}$ ]}\\
{\small \tt COHORT BY $\mathcal{L}$}
\end{tabbing}

In the above syntax, $e$ is the birth action that is specified by the data analyst
for the whole cohort query. The order of {\tt BIRTH FROM} and {\tt AGE ACTIVITIES IN}
clauses is irrelevant,
and the birth selection (i.e., $\sigma^b_{C,e}$) and age
selection (i.e., $\sigma^g_{C,e}$) clauses are optional.
We also introduce two keywords {\tt AGE} and {\tt COHORTSIZE} for data analysts
to retrieve the calculated columns produced by $\gamma^c_{\mathcal{L}, e, f_A}$
in the {\tt SELECT} list. Note that except for projection, we disallow other relational
operators such as $\sigma$ (i.e., SQL {\tt WHERE}) and $\gamma$ (i.e., SQL {\tt GROUP BY}),
and binary operators like intersection and join,
in a basic cohort query.

With the newly developed cohort operators, the cohort analysis task
presented in Example 1 can be expressed by the following query:


\begin{tabbing}
\small $Q_1$: {\tt SELECT country, COHORTSIZE, AGE, Sum(gold) as spent} \\
\quad\quad {\small \tt FROM $D$ AGE ACTIVITIES IN action = "shop"}\\

\quad\quad {\small \tt BIRTH FROM action = "launch" AND role = "dwarf"} \\
\quad\quad {\small \tt COHORT BY country}
\end{tabbing}

%

\subsection{Extensions}
\label{subsec:cohortquerytextension}

Our cohort query proposal can be extended in many directions to enable even more
complex and deep analysis. First, it would be great to mix cohort queries
with SQL queries in a single query. For example, one may want to use a SQL query to
retrieve specific cohort trends produced by a cohort query for further analysis.
If the cohort operation is specified by multiple attributes (e.g.,
{\tt COHORT BY time, country, role}), one may want to apply a SQL {\tt CUBE}
operator on top of the cohort query results for OLAP analysis.
This mixed querying requirement can be achieved by applying the standard SQL
{\tt WITH} clause to encapsulate a cohort query as a sub-query that can
be processed by an outer SQL query. The following example demonstrates
how to use a mixed query to retrieve specific cohort spent trends
reported by $Q_1$ for further analysis.

\begin{tabbing}
{\small \tt WITH cohorts AS ($Q_1$)}\\
{\small \tt SELECT cohort, AGE, spent FROM cohorts}\\
{\small \tt WHERE cohort IN ["Australia", "China"]}
\end{tabbing}

However, to ensure the correctness of a mixed query,
certain rules must be established to prevent SQL queries from
removing birth activity tuples incidently.
First, for a mixed query,
the outermost query must be a SQL query. That is, cohort queries can only
serve as sub-queries inside a mixed query.
Furthermore, we only allow SQL queries
(no matter sub-queries or the outer query) to reference
cohort sub-queries in the {\tt FROM} or {\tt WHERE} clauses and
prohibit a cohort sub-query from referencing other SQL or cohort sub-queries.
In addition, a mixed query should be evaluated in the order of first cohort
sub-queries and then others.
This ``cohort query first" evaluation scheme ensures that
no SQL queries can incidentally remove birth activity tuples
during query processing.

Another extension is to introduce binary cohort operators (e.g., join, intersection etc.)
for analyzing multiple activity tables. We leave the details of evaluating
a mixed query and other interesting extensions in a future paper.
In the rest of this paper, we shall focus on the approaches for evaluating
a single cohort query over a single activity table.

\subsection{Mapping Cohort Operations to SQL Statements}

Before leaving this section, we shall demonstrate that given
a materialized view (MV) built for a specific birth action,
the proposed cohort operators can be implemented by SQL sub-queries.
This enable us to pose cohort queries composed of
newly developed operators in the context of a non-intrusive
mechanism.


As shown in Section \ref{sec3}, the
MV approach stores each activity tuple of user $i$ along with
$i$'s birth attributes. Thus, to implement the birth selection operator,
one can use a SQL {\tt SELECT} statement
with a {\tt WHERE} clause specifying the birth selection condition
on the materialized birth attributes. Similarly, the age selection
operator can be simulated by a SQL {\tt SELECT} statement with
a {\tt WHERE} clause specifying the age selection condition
along with an additional predicate to include birth activity tuples.
The cohort aggregation operator can be implemented by
applying a SQL {\tt GROUP BY}
aggregation operation on the joined results between the
{\tt cohortSize} table and the qualified age activity tuples.


As an example, Figure~\ref{fig:mv:correspondence} demonstrates for $Q1$ of
Example 1 the correspondence between the three proposed cohort
operators and the equivalent SQL statements posed on the MV built for the {\tt launch}
birth action. As in
Figure~\ref{fig:sqlofq1}, the {\ttfamily player}, {\ttfamily action} and {\tt
time} attributes are respectively abbreviated to {\tt p}, {\tt a}, and {\tt t}.
{\tt bc, br, bt} and {\tt age} are four attributes additionally materialized along with
the original activity table. The first three attributes, {\tt bc, br} and {\tt bt},
respectively represent the birth attributes for {\tt country, role} and {\tt time}.
It should be noted that the SQL statements of Figure~\ref{fig:mv:correspondence}
are separated out for ease of exposition:
one can optimize them by combining Figure~\ref{subfig:mv:birth}
and~\ref{subfig:mv:age} into a single SQL statement, as we do in experiments.

\begin{figure*}[tbh]

	\centering
	\small
	\subfigure[$\sigma^b_{\text{role="dwarf", launch}}$]{
		\label{subfig:mv:birth}
		\begin{minipage}[t]{0.28\textwidth}
			\begin{tabbing}
				{\tt WITH birthView AS(} \\
				\quad {\tt SELECT p, a, t, gold,} \\
				\quad\quad {\tt bc, bt, age} \\
				\quad {\tt FROM $MV$}\\
				\quad {\tt WHERE br = "dwarf"}\\
				{\tt ),}\\
			\end{tabbing}
		\end{minipage}		
	}
	\subfigure[$\sigma^g_{{\text{action="shop", launch}}}$]{
		\label{subfig:mv:age}
		\begin{minipage}[t]{0.3\textwidth}
			\begin{tabbing}
				{\tt ageView AS (}\\
				\quad{\tt SELECT *}\\
				\quad{\tt FROM birthView}\\
				\quad{\tt WHERE a = "shop" OR}\\
				\quad\quad{\tt (t=bt AND a="launch")}\\
				{\tt ),}\\
			\end{tabbing}
		\end{minipage}
	}
		\subfigure[]{
			\label{subfig:mv:cs}
			\begin{minipage}[t]{0.3\textwidth}
				\begin{tabbing}
					{\tt cohortSize AS (}\\
					\quad{\tt SELECT bc as cohort,}\\
					\quad\quad{\tt Count(distinct p)}\\
					\quad\quad\quad{\tt as size}\\
					\quad{\tt FROM birthView}\\
					\quad{\tt GROUP BY bc),}\\
				\end{tabbing}
			\end{minipage}
		}
	\subfigure[$\gamma^c_{\text{country, launch, Sum(gold)}}$] {
		\label{subfig:mv:agg}
		\begin{minipage}[t]{0.45\textwidth}
			\begin{tabbing}
            \texttt{SELECT cohort, size, age,}\\
                \quad{\tt Sum(gold) as spent}\\
				\texttt{FROM ageView, cohortSize}\\
				\texttt{WHERE cohort = bc} \\
				\texttt{GROUP BY cohort, age}\\
				\\
			\end{tabbing}
		\end{minipage}
	}
		\caption{The correspondence between cohort operators and SQL statements of the MV approach}\label{fig:mv:correspondence}
	\end{figure*}

\section{COHANA: Cohort Query Engine}
\label{sec4}

\begin{figure}[tb]
  \centering
  \includegraphics[width=.28\textwidth]{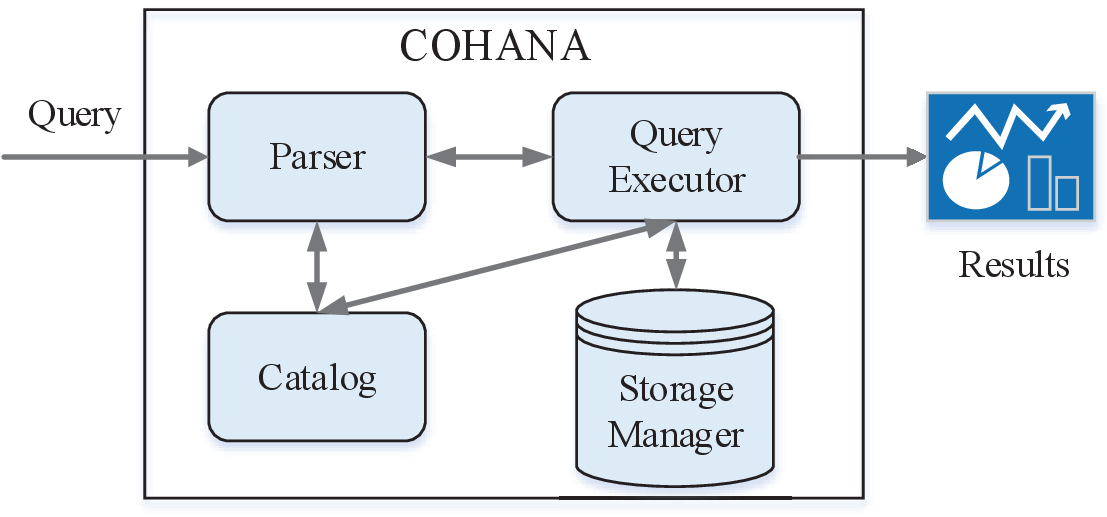}\\
  \caption{COHANA Architecture}\label{fig::cohana}
  \vspace{-5mm}
\end{figure}

To support cohort analytics with the newly designed cohort operators,
we present four extensions to a columnar database: 1) a fine tuned columnar storage
format for persisting activity tables; 2) a modified table scan operator
capable of skipping age activity tuples of unqualified users;
3) a native efficient implementation of cohort operators; 4)
a query planner capable of utilizing the cohort operator property
(i.e., Equation \eqref{cohortlaw}) for optimization.
We have implemented the proposed techniques in a columnar based
query engine, COHAHA, for performance study. Figure \ref{fig::cohana} presents
the architecture of COHAHA which includes four modules:
parser, catalog,  storage manager and query executor. The first two
modules are trivial, and we shall focus on the other two modules. 

\subsection{The Activity Table Storage Format}
\label{subsec:storage}


We store an activity table $D$ in the sorted order of
its primary key $(A_u, A_t, A_e)$. This storage layout
has two nice properties: 1) activity tuples of the same
user are clustered together; we refer to this as the
clustering property;
2) The activity tuples of each user
are stored in a chronological order; this is called
the time ordering
property. With these two properties, we can efficiently
find the birth activity tuple of any user for any birth action in a single
sequential scan. Suppose the activity tuples of user $i$ is stored between
$d_j$ and $d_k$. To find the birth activity tuple of $i$ for
any birth action $e$, we just iterate over each tuple between $d_j$ and
$d_k$ and return the first tuple $d_b$ satisfying $d_b[A_e] = e$.

We employ a chunking scheme and various
compression techniques to speed up cohort query processing.
We first horizontally partition the activity table into multiple data chunks
such that the activity tuples of
each user are included in exactly one chunk. Then, in each chunk, the activity
tuples are stored column by column.
For each column in a data chunk, we choose an appropriate compression scheme
for storing values based on the column type.

For the user column $A_u$, we choose Run-Length-Encoding (RLE) scheme.
The values in $A_u$ is stored as
a sequence of triples $(u, f, n)$, where $u$ is the user in $A_u$,
$f$ is the position of the first appearance of $u$ in the column,
and $n$ is the number of appearances of $u$ in the column.
We shall see in Section \ref{subsec:tablescan},
a modified table scan operator can directly
process these triples and efficiently skip to the activity tuples of the next user
if the birth activity tuple of the current user is not qualified with respect to
the birth selection condition.

For the action column $A_e$ and other string columns, we employ a two level
compression scheme presented in \cite{powerdrill} for storing
the values. More details of this encoding scheme can be found
in \cite{powerdrill}.
For each such column $A$, we first build and persist a global dictionary which
consists of the sorted unique values of $A$.
Each unique value of $A$ is then assigned a global-id, which is the position of
that value in the global dictionary.
For each data chunk, the sorted global-ids of the unique values of $A$ in that chunk
form a chunk dictionary. Given the chunk dictionary, each value of $A$ in that chunk
can be represented as a chunk-id, which is the position of the global-id of that
value in the chunk dictionary. The chunk-ids are then persisted immediately after the chunk
dictionary in the same order as the respective values appear in $A$.
This two level encoding scheme enables
the efficient pruning of chunks where no users perform the birth action.
For a given birth action $e$, we first perform a binary search
on the global index to find its global-id $g_i$. Then, for each data
chunk, we perform a binary search
for $g_i$ in the chunk dictionary. If $g_i$ is not found, we can safely
skip the current data chunk since no users in the data chunk perform $e$.

For $A_t$ and other integer columns, we employ a two-level delta encoding
scheme which is similar to the one designed for string columns. For each column
$A$ of this type, we first store the MIN and MAX value of $A$ for the whole
activity table as the global range. Then, for each data chunk,
the MIN and MAX values are extracted as the chunk range
from the segment of $A$ in that chunk and persisted as well.
Each value of the column segment is then finally stored as the delta (difference) between it
and the chunk MIN value.  Similar to the encoding scheme for string columns,
this two-level delta encoding scheme also enables the efficient pruning of chunks
where no activity tuples fall in the range specified in the birth selection
or age selection operation.

With the above two encoding schemes, the final representation of string columns
and integer columns are arrays of integers within a small range.
We therefore further employ
integer compression techniques to reduce the storage space. For each integer
array, we compute the minimum number of bits, denoted by $n$, to represent the
maximum value in the array, and then sequentially pack as many values as
possible into a computer word such that each value only occupies $n$ bits.
Finally, we persist the resulting computer words
to the storage device. This fixed-width encoding scheme is by no means
the most space-saving scheme. However, it enables the compressed values to be
randomly read without decompression. For each position in the original integer
array, one can easily locate the corresponding bits in the compressed computer
words and extract the value from these bits.
This feature is of vital importance for efficient cohort query processing.


\subsection{Cohort Query Evaluation}
\label{subsec:queryprocessing}

\begin{figure}[tb]
  \centering
  \includegraphics[width=.24\textwidth]{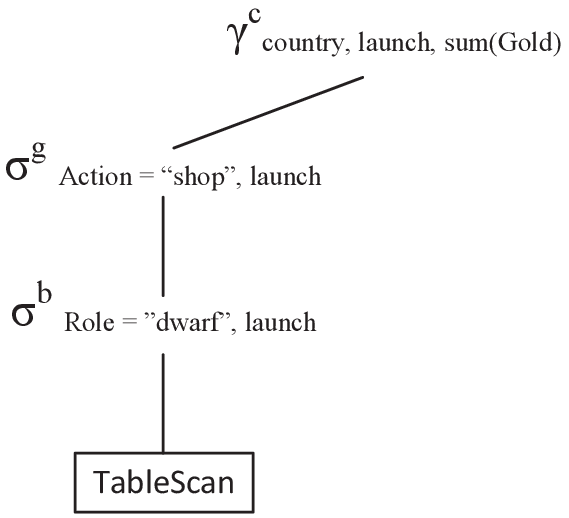}\\
  \caption{Query plan for $Q_1$}\label{fig1}
  \vspace{-3mm}
\end{figure}

This section presents how to evaluate a cohort query over the activity table
compressed with the techniques proposed in Section \ref{subsec:storage}.
We shall use the cohort query $Q_1$ as our running example. The overall
query processing strategy is as follows. We first generate a logical query plan,
and then optimize it by pushing down the birth selections along the plan.
Next, the optimized query plan is executed against each data chunk. Finally,
all partial results produced by the third step are merged together to produce
the final result. The final merging step is trivial and we shall only
present the first three steps.

The cohort query plan we introduced
in this paper is a tree of physical operators consisting of four operators:
{\tt TableScan}, birth selection $\sigma^b_{C,e}$, age selection $\sigma^g_{C,e}$
and cohort aggregation $\gamma^c_{\mathcal{L}, e, f_A}$.
Like other columnar databases, the projection operation is implemented in a pre-processing step:
we collect all required columns at query preparation stage and then pass those
columns to the {\tt TableScan} operator which retrieves the values for each
column.

In the query plan, the root and the only leaf node are the aggregation operator,
$\gamma^c_{\mathcal{L}, e, f_A}$, and the {\tt TableScan} operator,
respectively, and between them is a sequence of birth selection
operators and age selection operators.

Then, we push down the birth selection operators along the query plan
such that they are always below the age selection operators.
This push-down optimization is always feasible,
since according to equation \eqref{cohortlaw}, we can arbitrarily swap the order
of $\sigma^b_{C,e}$ and $\sigma^g_{C,e}$ operators in any sequence consisting of
these two operators. Figure \ref{fig1} shows the query
plan for the cohort query of $Q_1$.
We always employ this push-down optimization since,
as we shall see in Section \ref{subsec:tablescan}, a special designed
{\tt TableScan} implementation can efficiently skip age activity tuples
without further processing
for users whose birth activity tuples do not satisfy the birth selection
condition. Therefore, the cost of evaluating
birth selection operators before age selection operators is always less than
the cost incurred from the reverse evaluation sequence in terms of the number of
activity tuples processed.

After pushing down birth selections, the resulting query plan
will be executed against each data chunk. Before the execution, we apply an additional
filtering step by utilizing the $A_e$ column's two-level compression scheme
to skip data chunks where no users perform the birth action $e$. The concrete
processing strategy is presented in Section \ref{subsec:storage}.
In practice, we find that this intermediate filtering step is particularly
useful if the birth action is highly selective (i.e., only a few users
performed that birth action).

We will present the implementation of the physical operators in
the rest of this section.

\subsection{The {\tt TableScan} Operator}
\label{subsec:tablescan}

We augment the standard \texttt{TableScan} operator of columnar databases
for efficient cohort query processing.
The modified {\tt TableScan} operator performs scanning operation over
the compressed activity table that we proposed in Section \ref{subsec:storage}.
We mainly add two additional functions to a standard columnar database
{\tt TableScan} operator: \GetNextUser{} and \SkipCurUser{}. The \GetNextUser{}
function returns the activity tuple block of the next user; the \SkipCurUser{}
skips the activity tuples of the current user.

The modified {\tt TableScan} operator is implemented
as follows. For each data chunk, in the query initialization stage,
the {\tt TableScan} operator collects all (compressed) chunk columns referenced in the query
and maintains for each chunk column a file pointer which is initialized
to point to the beginning of that chunk column.
The implementation of \GetNext{} function
is identical to the standard {\tt TableScan} operator of a columnar database.

The \GetNextUser{} is implemented by first
retrieving the next triple $(u, f, n)$ of $A_u$ column and then advancing the file pointer
of each other referenced column to the beginning of the column segment corresponding to user $u$.
The \SkipCurUser{} is implemented in a similar way. When it is called, the \SkipCurUser{}
function first calculates the number of remaining
activity tuples of the current user,
and then advances the file pointers of all columns by the same
number.

\subsection{Cohort Algorithms}
\label{subsec:cohortalgo}

This section develops algorithms for the implementation of cohort operators
over the proposed storage format for activity tables.
\SetKw{KwOr}{OR}
\begin{algorithm}[tb]
\caption{$\sigma^b_{C,e}(D)$ operator implementation}\label{alg1}
\small
\Input{A data chunk $D$ and a birth action $e$}

\Fn{\GetBirthTuple{$d, e$}}{
  $i \leftarrow d[A_u]$ \;
  \While{$d[A_u] = i \wedge d[A_e]\neq e$}{
    $d \leftarrow D$.\GetNext{}\;
  }
  \Return $d$ \;
}
\BlankLine

\Fn{\Open{}}{
  $D$.\Open{} \;
  $u_c \leftarrow \emptyset$ \;
}
\BlankLine
\Fn{\GetNext{}}{
  \If{$u_c$ has more activity tuples} {
    \Return $D.\GetNext{}$
  }

  \While{there are more users in the data chunk} {
    $(u, f, n) \leftarrow D.\GetNextUser{}$\;
    $u_c \leftarrow u$\;
    $d \leftarrow D.\GetNext{}$ \;
    $d^b \leftarrow \GetBirthTuple{$d, e$}$ \;
    {\tt Found} $\leftarrow C(d^b)$ \;
    \If{\tt Found} {
      \Return $d$\;
    }
    $D.\SkipCurUser{}$\;
  }
}
\end{algorithm}

Algorithm~\ref{alg1} presents the implementation of the birth selection
operator $\sigma^b_{C,e}$.
It employs an auxiliary function\\
\GetBirthTuple{$d, e$} (line 1 -- line 5) for finding the birth
activity tuple
of user $i = d[A_u]$, given that $d$ is the first activity
tuple of $i$ in the data chunk and $e$ is the birth action.
The \GetBirthTuple{} function finds $i$'s birth activity
tuple by iterating over each next tuple $d \in D$
and checks whether $d$ belongs to $i$ and whether
$d[A_e]$ is the birth action $e$ (line 3).
The first activity tuple $d$ matching the
condition is the required birth activity tuple.

To evaluate $\sigma^b_{C,e}$, Algorithm \ref{alg1}
first opens the input data chunk $D$
and initializes the global variable $u_c$ (line 7 -- line 8)
which points to the user currently being processed.
In the \GetNext{} function, we return
the next activity tuple $d$ of $u_c$ if $u_c$ is
qualified with respect to the birth selection condition (line 11).
If $u_c$'s activity tuples are exhausted, we retrieve
the next user block by calling the \GetNextUser{} function of
the {\tt TableScan} operator (line 13).
Then, we find the birth activity
tuple of the new user and check if it satisfies the birth selection condition
(line 16 -- line 17). If the new user is qualified, its birth activity tuple
will be returned; otherwise all the activity tuples of this user will be skipped
using the \SkipCurUser{} function so that its next user can be ready for
processing. Therefore, one can continuously call the \GetNext{} function to
retrieve the activity tuples of users that are qualified with respect to the
birth selection condition.

The implementation of $\sigma^g_{C,e}$ is much simpler than
$\sigma^b_{C,e}$. We also employ the user block processing
strategy. For each user block, we first locate the birth activity tuple
and then return the birth activity tuple and qualified
age activity tuples.

\begin{algorithm}[tb]
\caption{$\gamma_{\mathcal{L},e,f_A}(D)$ operator implementation}\label{alg3}
\small
\Input{A data chunk $D$, a birth action $e$, an attribute list $\mathcal{L}$}

\Fn{\Open{}} {
  $D.\Open{}$ \;
  $H^c \leftarrow \emptyset$ \tcp*[f]{Cohort size hash table}\;
  $H^g \leftarrow \emptyset$ \tcp*[f]{Cohort metric hash table}\;
  \While{there are more users in $D$} {
    $(u, f, n) \leftarrow D.\GetNextUser{}$ \;
    $u_c \leftarrow u$ \;
    $d \leftarrow D.\GetNext{}$ \;
    $d^b \leftarrow D.\GetBirthTuple{$d, e$}$ \;
    \If{$u_c$ is qualified} {
      $H^c[d^b[\mathcal{L}]]++$ \;
      \While{$u_c$ has more qualified age activity tuples} {
        $g \leftarrow d[A_t] - d^b[A_t] $ \;
        update $H^g[d^b[\mathcal{L}]][g]$ with $f_A(d)$ \;
      }
    }
  }
}
\BlankLine

\Fn{\GetNext{}} {
  Retrieve next key $(c, g)$ from $H^g$ \;
  \Return $(c, g, H^c[c], H^g[c][g])$
}
\end{algorithm}

Algorithm \ref{alg3} presents the implementation of $\gamma^c_{\mathcal{L}, e, f_A}$
operator. The main logic is implemented in the \Open{} function.
The function first initializes two hash tables $H^c$ and $H^g$ which
respectively store the cohort size and per data chunk
aggregation result for each (cohort, age) partition (line 2 -- line 6).
Then, the \Open{} function iterates over each user block and updates $H^c$
for each qualified user (determined by $\sigma^b_{C,e})$ and
$H^g$ for all qualified age activity tuples (determined by $\sigma^g_{C,e}$)
(line 10 -- line 14). To speed up the query processing, we further
follow the suggestions presented in \cite{datacube, powerdrill} and
use array based hash tables for aggregation. In practice, we find
that the use of
array-based hash tables in the inner loop of cohort aggregation
significantly improves the performance since modern CPUs
can highly pipeline array operations.

\subsection{Optimizing for User Retention Analysis}
\label{subsec:otheropt}


One popular application of cohort analysis is to show the trend of
user retention \cite{mixpanel}. These cohort queries
involve counting distinct number of users for each (cohort, age)
combination. This computation is very costly in terms of memory
for fields with a large cardinality, such as $A_u$.
Fortunately, our proposed storage format has a nice property
that the activity tuples of any user are included in only one chunk.
We therefore implement a \texttt{UserCount()} aggregation function for
the efficient counting of distinct users by performing counting against each chunk
and returning the sum of the obtained numbers as the final result.

\subsection{Analysis of Query Performance}
\label{subsec:algorithmanalysis}

Given there are $n$ users in the activity table $D$, each user
produces $m$ activity tuples, it can be clearly seen that, to evaluate
a cohort query composed of $\sigma^b_{C,e}$, $\sigma^g_{C,e}$ and
$\gamma^c_{\mathcal{L}, e, f_A}$ operators, the
query evaluation scheme we presented so far only needs to process
$O(l\times m)$ activity tuples in a single pass,
where $l$ is the number of qualified users
with respect to the birth selection condition.
Therefore, the query processing time grows linearly with $l$,
and therefore approaches optimal performance.

\section{A Performance Study}
\label{sec5}

This section presents a performance study to evaluate the effectiveness
of our proposed COHANA engine. We mainly
perform two sets of experiments.
First, we study the effectiveness of COHANA,
and its optimization techniques.
In the second set of experiments,
we compare the performance
of different query evaluation schemes. We implement the SQL based approach and the materialized
view based approach on two relational databases: Postgres and MonetDB, and compare
the performance of these two systems with COHANA. For the two relational
databases, we allow them to use all the free memory for buffering, and leave all other
settings as default.

\subsection{Experimental Environment}

All experiments are run on a high-end workstation. The workstation is equipped with
a quad-core Intel Xeon E3-1220 v3 3.10GHz processor and 8GB of memory. The disk speed
reported by \texttt{hdparm} is 14.8GB/s for cached reads and 138MB/s for buffered
reads.

The dataset we used is produced by a real mobile game application.
The dataset consists of 30M activity tuples contributed by 57,077 users
worldwide from 2013-5-19 to 2013-06-26, and occupies a disk space of 3.6GB in
its raw csv format. In addition to the required
user, action and action time attributes, we also include the country, city and role as
dimensions and session length and gold as measures.
Users in the game played 16 actions in total, and we choose the {\tt launch},
{\tt shop} and {\tt achievement} actions as the birth actions.
In addition, we manually scale the dataset and study
the performance of three cohort query evaluation schemes on different dataset size.
Given a scale factor X, we produce a dataset consisting of X times users.
Each user has the same activity tuples as the original dataset
except with a different user attribute.

For the SQL based approach, we manually translate the cohort
query into a SQL query as exemplified in Figure~\ref{fig:sqlofq1}.
For the materialized view approach,
we manually materialize the view using \texttt{CREATE TABLE AS} command.
For each birth action, we materialize the age and a birth attribute
set of time, role, country and city attribute in its materialized view.
This materialization scheme adds 15 additional columns to the original table
by performing six joins in total.
We also build a cluster index on the primary key and indices on birth attributes.
For COHANA, we choose the chunk size to be 256K.

\subsection{Benchmark Queries}
\label{benchquery}
We design four queries (described with COHANA's cohort query syntax)
for the benchmark by incrementally adding the cohort operators we proposed in this paper.
The first query Q1 evaluates a single cohort aggregation operator. The second query
Q2 evaluates a combination of birth selection and cohort aggregation. The third
query Q3 evaluates a combination of age selection
and cohort aggregation. The fourth query Q4 evaluates a combination of all three
cohort operators.  For each query, we report the average execution time of five runs for each system.

Q1: For each country launch cohort, report the number of retained users who did at least
one action since they first launched the game.
\begin{tabbing}
\small\texttt{SELECT country, CohortSize, Age, UserCount()}\\
\small\texttt{FROM GameActions BIRTH FROM action = "launch"} \\
\small\texttt{COHORT BY country}
\end{tabbing}

Q2: For each country launch cohort born in a specific date range, report the number of retained
users who did at least one action since they first launched the game.
\begin{tabbing}
\small\texttt{SELECT country, COHORTSIZE, AGE, UserCount()}\\
\small\texttt{FROM GameActions BIRTH FROM action = "launch" AND} \\
\small\quad\quad\texttt{time BETWEEN "2013-05-21" AND "2013-05-27"} \\
\small\texttt{COHORT BY country}
\end{tabbing}

Q3: For each country shop cohort, report the average gold they spent in shopping
since they made first shop in the game.
\begin{tabbing}
\small
\small\texttt{SELECT country, COHORTSIZE, AGE, Avg(gold)}\\
\small\texttt{FROM GameActions BIRTH FROM action = "shop"} \\
\small\texttt{AGE ACTIVITIES IN action = "shop"} \\
\small\texttt{COHORT BY country}
\end{tabbing}

Q4: For each country shop cohort, report the average gold they spent
in shopping in their birth country where they
were born with respect to the dwarf role in a given date range.
\begin{tabbing}
\small\texttt{SELECT country, COHORTSIZE, AGE, Avg(gold)}\\
\small\texttt{FROM GameActions BIRTH FROM action = "shop" AND} \\
\small\quad\quad\texttt{time BETWEEN "2013-05-21" AND "2013-05-27" AND} \\
\small\quad\quad\texttt{role = "dwarf" AND} \\
\small\quad\quad\texttt{country IN ["China", "Australia", "United States"]} \\
\small\texttt{\small AGE ACTIVITIES IN action$\,$=$\,$"shop" AND country$\,$=$\,$Birth(country)} \\
\small\texttt{COHORT BY country}
\end{tabbing}

In order to investigate the impact of the birth selection operator and the age
selection operator on the query performance of COHANA, we further design two
variants of Q1 and Q3 by adding to them a birth selection condition
(resulting in Q5 and Q6) or an age selection condition (resulting in Q7 and Q8).
The details of Q5-Q8 are shown below.

Q5: For each country launch cohort, report the number of retained users who did at least
one action during the date range [d1; d2] since they first launched the game.
\begin{tabbing}
\small\texttt{SELECT country, COHORTSIZE, AGE, UserCount()}\\
\small\texttt{FROM GameActions} \\
\small\texttt{BIRTH FROM action = "launch" AND time BETWEEN $d_1$ AND $d_2$} \\
\small\texttt{COHORT BY country}
\end{tabbing}

Q6: For each country shop cohort, report the average gold they spent in shopping
during the date range [d1; d2] since they made their first shop in the game.
\begin{tabbing}
\small
\texttt{SELECT country, COHORTSIZE, AGE, Avg(gold)}\\
\small\texttt{FROM GameActions} \\
\small\texttt{BIRTH FROM action = "shop" AND time BETWEEN $d_1$ AND $d_2$} \\
\small\texttt{AGE ACTIVITIES IN action = "shop"} \\
\small\texttt{COHORT BY country}
\end{tabbing}

Q7: For each country launch cohort whose age is less than $g$, report the number of retained
users who did at least one action since they first launched the game.
\begin{tabbing}
\small
\small\texttt{SELECT country, COHORTSIZE, AGE, UserCount()}\\
\small\texttt{FROM GameActions BIRTH FROM action = "launch"}\\
\small\texttt{AGE ACTIVITIES in AGE < $g$} \\
\small\texttt{COHORT BY country}
\end{tabbing}

Q8: For each country shop cohort whose age is less than $g$, report the average gold they
spent in shopping since they made their first shop in the game.
\begin{tabbing}
\small\texttt{\small SELECT country, COHORTSIZE, AGE, Avg(gold)}\\
\small\texttt{\small FROM GameActions BIRTH FROM action = "shop"} \\
\small\texttt{AGE ACTIVITIES IN action = "shop" AND AGE < $g$} \\
\small\texttt{COHORT BY country}
\end{tabbing}

\subsection{Performance Study of COHANA}

In this section we report on a set of experiments in which we vary chunk
size and birth/age selection condition and investigate
how COHANA adapts to such variation.

\subsubsection{Effect of Chunk Size}

\begin{figure*}[tbp]
	\centering
	\subfigure[Q1] {
		\includegraphics[width=0.259\textwidth]{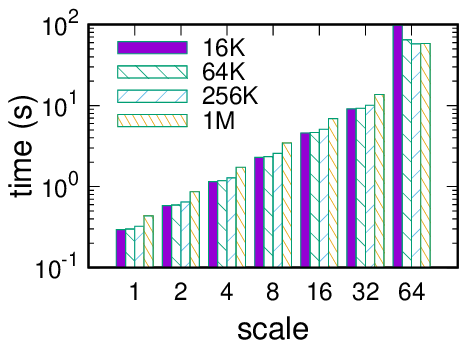}
		\label{subfig::csize::q1}
	}
    \hspace{-20pt}
	\subfigure[Q2] {
		\includegraphics[width=0.259\textwidth]{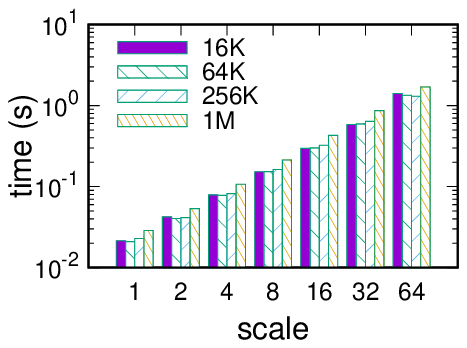}
		\label{subfig::csize::q2}
	}
    \hspace{-20pt}
	\subfigure[Q3] {
		\includegraphics[width=0.259\textwidth]{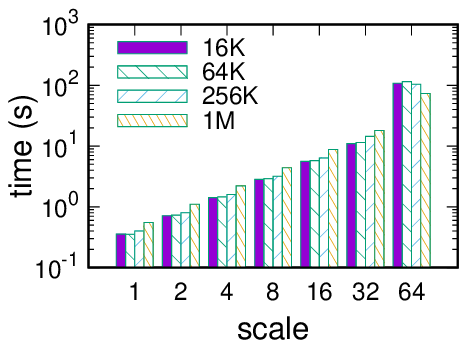}
		\label{subfig::csize::q3}
	}
    \hspace{-20pt}
	\subfigure[Q4] {
		\includegraphics[width=0.259\textwidth]{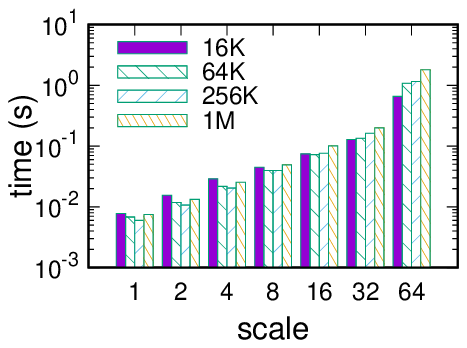}
		\label{subfig::csize::q4}
	}
	\caption{COHANA's performance under varying chunk size}
	\label{fig::csize}
\end{figure*}

\begin{figure*}[tb]
\begin{minipage}[t]{.25\linewidth}
  \centering
  \includegraphics[width=1.0\textwidth]{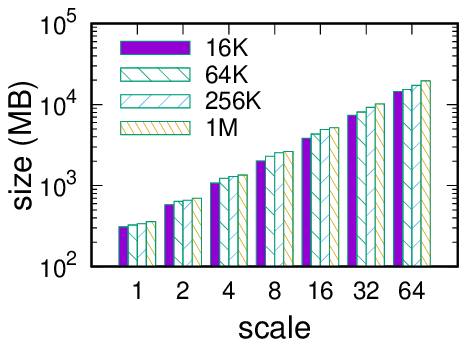}\\
  \caption{Effect of chunk size on storage space}\label{fig::space}
\end{minipage}
    \hspace{-5pt}
\begin{minipage}[t]{.25\linewidth}
  \centering
  \includegraphics[width=\textwidth]{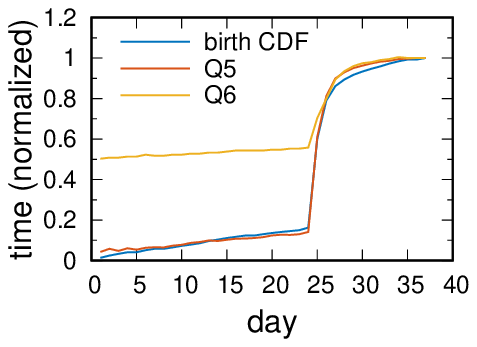}\\
  \caption{Effect of birth selection}\label{fig::bs}
\end{minipage}
    \hspace{-5pt}
\begin{minipage}[t]{.25\linewidth}
  \centering
  \includegraphics[width=1\textwidth]{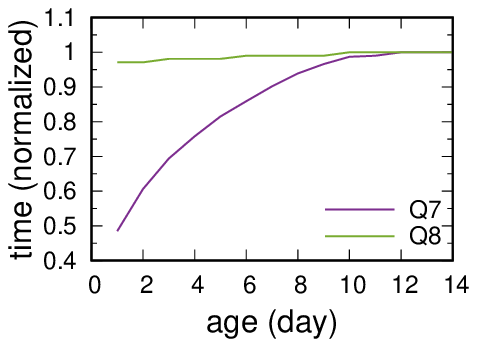}\\
  \caption{Effect of age selection\quad\quad}\label{fig::age}
\end{minipage}
\hspace{-5pt}
\begin{minipage}[t]{.25\linewidth}
  \centering
  \includegraphics[width=1.0\textwidth]{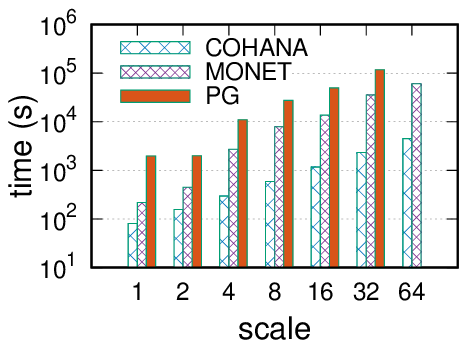}\\
  \caption{Time for generating MV}\label{fig::tsf}
\end{minipage}
\end{figure*}

Figures \ref{fig::csize} and \ref{fig::space} respectively present the storage space COHANA requires for
the activity table compressed with different chunk sizes,
and the corresponding query performance.
It is clearly seen from Figure \ref{fig::space} that increasing the chunk size also
augments storage cost.
This is because an increase in the size of a chunk will lead to more
players included
in that chunk. As a result, the number of distinct values in the columns of each chunk also increases,
which in turn requires more bits for encoding values.
We also observe that cohort queries can be processed slightly faster
under a smaller chunk size than a larger one. This is expected
as fewer bytes are read. However, for large datasets, a
larger chunk size can be a better choice.
For example, at scale 64, COHANA processes Q1 and Q3 most efficiently under 1M
chunk size. This is because the processing of Q1 and Q3 at scale 64 is dominated
by disk accesses, whose granularity is normally a 4KB block. Compared with a
large chunk size, a small one leads to more part of the neighbouring columns to
be simultaneously read when reading a compressed chunk column, and hence results
in a longer disk read time and a lower memory efficiency due to the memory
contention
between the useful columns and their unused neighbours within the same chunk.

\subsubsection{Effect of Birth Selection}

In Section~\ref{subsec:algorithmanalysis}, we claim that the running time of COHANA is bounded
by $O(n)$ where $n$ is the total number of qualified users. This experiment studies
the query performance of COHANA with respect to the birth selection selectivity.
We run Q5 and Q6, which are respectively a variant of Q1 and Q3, by fixing $d1$ to be the
earliest birth day, and incrementing $d2$ by one day each time. The dataset used
in this experiment is at scale 1.

Figure \ref{fig::bs} presents the processing times of Q5 and Q6
which are respectively normalized by
that of Q1 and Q3. The cumulative distribution of user births is also given
in this figure. We do not differentiate the birth distributions between the
birth actions of launch and shop, as the birth distributions with respect to
both birth actions are similar.
It can be clearly observed from this figure that the processing time of Q5
highly coincides with the birth distribution.
We attribute this coincidence to the optimization of
pushing down the birth selection operator
and the refined birth selection algorithm which is capable of
skipping unqualified users. The processing time of Q6, however, is not very
sensitive to the birth distribution. This is because in Q6, users are born
with respect to the shop action, and there is a cost in finding the
birth activity tuple for each user. This cost is avoided in Q5 as the first activity tuple of each
user is the birth activity tuple of this user (recall that the first action each user performed is
launch).

\subsubsection{Effect of Age Selection}
In this experiment, we run Q7 and Q8, another variant of Q1 and Q3,
on the dataset of scale 1 by varying $g$ from 1 day to 14 days to
study the query performance of COHANA under different age selection conditions.
Figure~\ref{fig::age} presents the result of this experiment. As in
Figure~\ref{fig::bs}, the processing times of Q7 and Q8 are also respectively
normalized by that of Q1 and Q3.  It can be seen from
this figure that the processing times of Q7 and Q8 exhibit different trends.
Specifically, the processing time of Q7 increases almost linearly, while
the processing time for Q8 increases slowly. The reason for this difference is
that the performance of Q7 is bounded by the number of distinct users within the
given age range, which grows almost linearly with age range.
For Q8, the processing time mainly depends on finding the birth activity tuples and
the aggregation performed upon the shop activity tuples.
The cost of the former operation is fixed across various age ranges, and
the cost of the latter operation does not change dramatically as
the number of shop activity tuples grows slowly with the age -- the aging
effect.

\subsection{Comparative Study}

\begin{figure*}[tbp]
	\centering
        \includegraphics[width=1\textwidth]{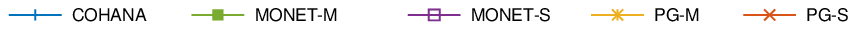}\\
	    \vspace{-5pt}
	\subfigure[Q1] {
		\includegraphics[width=0.259\textwidth]{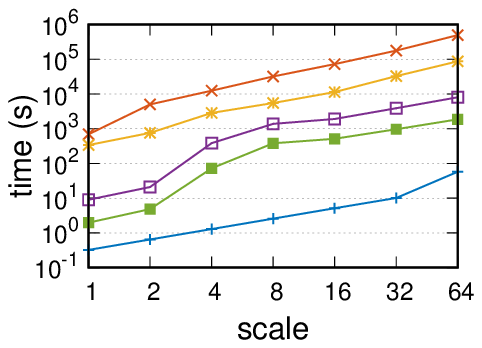}
		\label{subfig::cmp::q1}
	}
    \hspace{-20pt}
	\subfigure[Q2] {
		\includegraphics[width=0.259\textwidth]{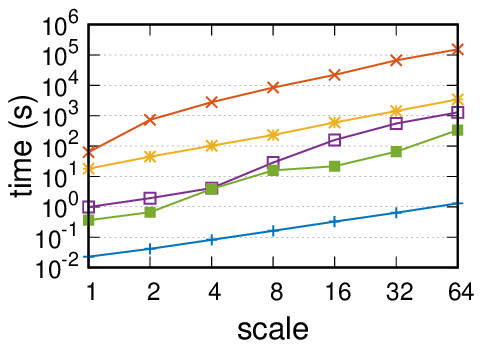}
		\label{subfig::cmp::q2}
	}
    \hspace{-20pt}
	\subfigure[Q3] {
		\includegraphics[width=0.259\textwidth]{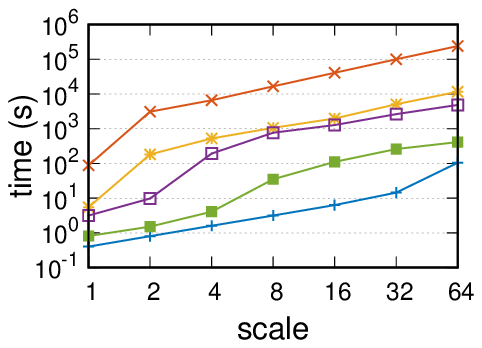}
		\label{subfig::cmp::q3}
	}
    \hspace{-20pt}
	\subfigure[Q4] {
		\includegraphics[width=0.259\textwidth]{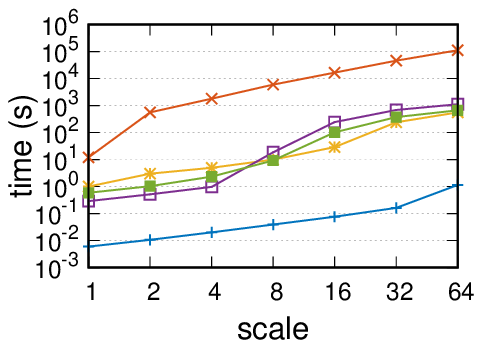}
		\label{subfig::cmp::q4}
	}
	\caption{Performance comparison among different evaluation schemes}
	\label{fig::cmp}
\end{figure*}

Figure \ref{fig::cmp} reports for each scale (factor) the execution time that each system takes to
execute the four queries. The results of the Postgres and the MonetDB databases
are respectively shown in the lines labelled by ``PG-S/M" and in those labelled
by ``MONET-S/M", where ``S" and ``M" respectively mean the SQL
and the materialized view approaches.
As expected, the SQL based approach is the slowest as it needs multiple joins
for processing cohort queries. With the elimination of joins, the materialized
view based approach can reduce the query processing time by an order of
magnitude. This figure also shows the power of columnar storage in terms of
cohort query processing. MonetDB, a state-of-the-art columnar database, can be
up to two orders faster than Postgres.

Although the combination of a materialized view and columnar storage can address
cohort queries reasonably well on small datasets; however, it is not able to
handle large datasets. For example, it takes
half an hour to process Q1 at scale 64.
The proposed system, COHANA, is able to perform extremely well not only on
small datasets, but also on large datasets. Moreover, for each query, COHANA is able to
perform better than the MonetDB equipped with the materialized view
at any scale. The performance gap between them is one to two orders of magnitude in most
cases, and can be up to three orders of magnitude (Q4 at scale 32). We also
observe that the two retention queries (Q1 and Q2) enjoy a larger
performance gain than Q3 does and in part attribute it to the
optimization Section~\ref{subsec:otheropt} presents for user retention analysis.
Finally, the generation of the materialized view is much more expensive than COHANA. As
shown in Figure~\ref{fig::tsf}, at scale 64, MonetDB needs more than 60,000 seconds (16.7
hours) to generate the materialized view from the original activity table.
This time cost is even more expensive in Postgres, which needs more than 100,000
seconds (27.8 hours) at scale 32. The result for Postgres at scale 64 is not
available as Postgres is not able to generate the materialized
view before using up all free disk space, which also implies a high storage cost
during the generation of the materialized view. 
In a sharp contrast, COHANA only needs 1.25
hours to compress the activity table of scale 64.

\section{Related Work}
\label{sec6}

The work related to ours is the database support for data analysis and cohort analysis.
The requirement to support data analysis inside a database system has a long history. The early
effort is the SQL {\tt GROUP BY} operator and aggregate functions.
These ideas are generalized with the {\tt CUBE} operator \cite{datacube}.
Traditional row-oriented databases are inefficient for {\tt CUBE} style OLAP analysis.
Hence, columnar databases are proposed for solving the efficiency issue
\cite{monetdbx100,monetdb,cstore}. Techniques such as
data compression \cite{compressdatabase,datacompression}, query processing on compressed data
\cite{abadi06,bitmapcompress,bitweaving}, array based
aggregation \cite{multiaggregates,array}, and materialized view based approaches \cite{mv} are
proposed for speeding up OLAP queries.
Albeit targeting OLAP queries which are defined on relational operators that
are generally not applicable to cohort queries, the above techniques can also be used to
accelerate the processing of cohort queries as we have shown in Section~\ref{sec4}.

Cohort analysis originates from social science \cite{cohortbook}.
However, the cohort analysis approach presented
in social science literatures has two limitations:
1) lack of a way for specifying a subset of users
or activity tuples for analysis; 2) only use time attribute to identify
cohorts. These two limitations are recognized in modern
analytical software package \cite{googlecohort, mixpanel, rjmetrics}.
These software somehow try to solve these limitations in
their respective application domains. For example, MixPanel
allows data analysts to select user segment for cohort analysis.
But none of the solutions we investigated so far is general.
For example, none of the software support {\tt Birth()} filtering
we present in this paper.
An implicit cohort analysis is conducted in \cite{cai2015multi} to study the behavior of
users in a private BitTorrent community using the SQL approach.
Compared to the above works, our effort not only generalizes the cohort analysis
for broader spectrum of applications,
but also is the first attempt to extend database systems to support the generalized cohort analysis.

\section{Conclusions}
\label{sec7}

Cohort analysis is a powerful tool for
finding unusual user behavioral trends
in large activity tables. This paper has conducted
the first investigation of database support for cohort analysis.
Consequently, we have introduced an extended
relation for modeling activity data and
extended SQL with three new operators for composing cohort queries.
We have developed a columnar based query engine, COHAHA,
for efficient cohort query processing.
Our experimental results showed that COHANA
can achieve two orders faster query performance than simply running
SQL queries over conventional database systems,
demonstrating the possible benefit of
extending a database system for cohort queries over implementing
cohort queries on top of it.


%

\balance

\bibliographystyle{abbrv}
\bibliography{sigproc}  

\end{document}